\documentclass{aastex}
\usepackage{spr-astr-addons}
\usepackage{url}
\usepackage{graphicx}
\usepackage{natbib}
\usepackage{amsmath}

\RequirePackage{color}

\begin{document}

\title{``Anti-glitches" in the Quark-Nova model for AXPs I}

\author{R.  Ouyed} \and \author{D.  Leahy} \and \author{N.  Koning}
\affil{Department of Physics and Astronomy, University of Calgary, Calgary, AB, Canada, T2N 1N4}

\begin{abstract}

In the Quark-Nova model, Anomalous X-ray Pulsars (AXPs) are quark stars surrounded by a degenerate iron-rich Keplerian ring (a few stellar radii away).
AXP bursts are caused by accretion of chunks from the inner edge of the ring following magnetic field
penetration.  For bright bursts,
 the inner disk is prone to radiation induced warping which can tilt it into counter-rotation (i.e. retrograde).
  For AXP 1E2259$+$586, the 2002 burst satisfies the condition for  the formation of a retrograde inner ring. We hypothesize
   the 2002  burst reversed the inner ring  setting the scene for the 2012 outburst and  ``anti-glitch" when 
    the retrograde inner ring was 
    suddenly  accreted  leading to the basic  observed properties of the 2012 event.

\end{abstract}


\section{Introduction}

Archibald et al. (2013) recently reported the first  observation of a sudden spin-down  in an AXP (1E 2259$+$586).
 This  spin-down was associated with 
an X-ray burst  with peak luminosity of the order of  $10^{38}$ erg s$^{-1}$  that lasted for about 
36 ms.  This puzzling ``anti-glitch"   begs for an explanation (e.g. Lyutikov 2013; Tong 2014; Katz 2014).

 In the Quark-Nova model (Ouyed et al. 2002; Ker\"anen et al. 2005; Ouyed\&Leahy 2009; Niebergal et al. 2010; Ouyed et al. 2013), Soft Gamma-ray Repeaters (SGRs) and AXPs  undergo outbursts from accretion. 
For SGRs,  accretion is from a co-rotating shell (see Ouyed et al. 2007a; hereafter  OLNI) whereas AXPs accrete from a Keplerian
  ring (see Ouyed et al. 2007b; hereafter  OLNII)\footnote{Other fall-back accretion models of SGRs/AXPs have been proposed, and may account for the anti-glitch as in the Quark-Nova model (e.g. Katz 1994; Alpar 2001; Tr\"umper et al. 2010).  However, the Keplerian  ring in our model is unlike a fall-back disk around a neutron star. The ring is rich in heavy elements, is very close to the quark star (a few stellar radii away) and is degenerate. Similar ring formation when a neutron star is born appears implausible since the proto-neutron star  is too large. In addition there is a strong outflow of energy during or immediately following the SN which may 
 also prohibited the formation of a degenerate Keplerian ring.}.  In both cases, the shell/ring is located a few stellar radii away from the
  quark star and consist of degenerate iron-rich material from the ejected neutron star crust.  In our model, anti-glitches are expected in SGRs (see  section 6.2 in OLNI).  Here we focus
  on AXPs and  refer the interested reader to OLNII, Ouyed et al. (2010; hereafter OLNIV) 
  and Ouyed et al. (2011; hereafter OLNV) for more details on the AXPs in our model.

   The Quark-Nova model for AXPs  normally leads to sudden spin-up during bursts, caused
  by accretion from the inner Keplerian ring. However, as we show below it also allows 
  for sudden spin-down during bursts if a preceding burst is bright enough to reverse the inner ring. 
  Here we show  that the warping instability (Pringle 1992\&1996) could develop following a
 bright burst which leads to a retrograde inner ring. 
  The possibility of  spin-down torques  induced by  a retrograde Keplerian accretion disk  has previously been
explored (e.g. Nelson et al. 1997).   Such torques could also occur if warping leads to 
an  inner disk  tilted by more than 90 degrees (e.g. van Kerkwijk et al. 1998).

\section{Our model}

\subsection{Ring properties}
\label{sec:ring}

The  general properties of the Keplerian ring (and other general features of our model)  can be found in \S 2 in OLNII and \S 2 in OLNIV.  We define the  ring's  inner radius as $R_{\rm in}$ and the outer radius as $R_{\rm out}$ (see Fig. 1 in OLNV
 for an illustration of the ring's structure, geometry and the star-ring configuration; see also figure Fig. 2 in OLNIV).
Viscosity causes the ring's outer radius to spread radially outwards 
at a rate given by  (see Appendix A in OLNII) 
 $R_{\rm out}\sim 105\ {\rm km} ~ (T_{\rm keV, 0.2}^{\rm q})^{5/4} t_{\rm yr, 10^4}^{1/2}$ 
 with  $t_{\rm yr, 10^4}$ being the age of the system in units of $10^4$ years while the ring's temperature during
 quiescence (superscript ``q") is given  in units of 0.2 keV.  For AXP 1E2259$+$586 with an estimated age of $\sim  2.5\times 10^4$ years (see Tendulkar et al. 2013) 
 this gives $R_{\rm out}\sim$ 166 km; we thus use 200 km as our fiducial value for $R_{\rm out}$.
 Since the source spends most of its life in quiescence, $R_{\rm out}$ spreading 
  is mainly determined by the ring's temperature during this phase;  $T_{\rm ring}^{\rm q}$ is
  in the sub-keV range while the temperature during burst, $T_{\rm ring}^{\rm b}$,  is in the keV range.

The ring's average density is found from  equation $\rho_{\rm ring}\simeq m_{\rm ring}/(2 H_{\rm out} \pi R_{\rm out}^2)$.
Here, $H_{\rm out}\sim 54 \  {\rm km}\ \rho_{\rm ring, 3}^{1/6}R_{\rm out, 200}^{3/2}$ is   the ring's vertical scale-height at $R_{\rm out}$ with the ring's average density\footnote{The ring's density is above the  non-degeneracy (nd) limit.  
 The maximum density of the ring below which the matter is non-degenerate is found by
setting the ring's temperature equal to the Fermi temperature, $T_{\rm ring}  = T_{\rm Fermi}$, which gives $\rho_{\rm ring, nd} <  100\ {\rm g\ cm}^{-3}$
 at sub-keV temperatures (see Appendix A in OLNII)}, $\rho_{\rm ring, 3}$,  given in units of $10^3$ g cm$^{-3}$ and
 the ring's outer radius $R_{\rm out, 200}$ given in units
of 200 km (in old sources $R_{\rm out} >> R_{\rm in}$).  This gives 
   $\rho_{\rm ring}\sim  1.1\times 10^3 \ {\rm g\ cm}^{-3}\ \times m_{\rm ring, -8}^{6/7}/R_{\rm out, 200}^3$
  with $m_{\rm ring, -8}$ being the mass of
the ring in units of $10^{-8}M_{\odot}$.  We chose   $m_{\rm ring}= 10^{-8}M_{\odot}$ 
as our fiducial value to ensure that for a QS with a birth period exceeding  $\sim 5$ ms 
 the fall-back material has enough angular momentum to form a Keplerian ring (see
  eq. 8 in OLNI and \S 2 in OLNII for more  details).  A value   $m_{\rm ring}= 10^{-8}M_{\odot}$ also  offers the advantage that the ring is easier to reverse  for typical burst energies in the QN model.

 The iron-rich ring is made  of ions in the strong-coupling regime which solidifies when the Coulomb parameter is  $\Gamma \sim 175$ (Nagara et al. 1987). The solidification temperature is estimated at  $T_{\rm s}\sim  9.5\ {\rm keV}\times  (\frac{\rho}{10^8\ {g\ cm}^{-3}})^{1/3}$ (De Blasio  1995;  Baiko \& Yakovlev 1995;  Pothekin 1999), so that here with  $\rho_{\rm ring} \sim 10^3$ g cm$^{-3}$, we get $T_{\rm s} \sim 0.2$ keV, i.e the ring is solid except perhaps during bursts.
  The ring  made of cold (sub-keV) solid matter  is thus prone to tidal shearing.
As shown in \S 2.1 in OLNII, the ring fractures   into hundreds of  cylinders (which we dubbed ÒwallsÓ) each of thickness  $\Delta r_{\rm wall}\sim 862.5\ {\rm cm} f_{\rm Fe} R_{\rm in, 25}^{3/2}$;
 here the ring's inner radius, $R_{\rm in}$ is given in units of 25 km and
 $f_{\rm Fe}$ a dimensionless tensile strength of the ring's material. 
 The walls are stacked against each other but separated by  a degenerate fluid;
effectively  each wall  in high pressure contact
with the next through a melted (noncrystalline) material.
  The typical mass of a wall is of the order of $M_{\rm wall}\sim 10^{-11}M_{\odot}$ in our model.

 \subsection{Ring warping and tilting conditions: the 2002 burst}
 
The condition for  warping was presented in Pringle (1992 and 1996) for a geometrically 
 thin and optically thick accretion disk.  
   Numerical simulations by  van Kerkwijk et al (1998; their Fig. 1) show the
tilt oscillating between  90 and 180 degrees. Wijers \& Pringle (1999) simulations, in a different regime,  can reproduce
 the observed tilt of  Her X-1 of about 25 degrees;  the inner disc  in such systems was shown to tilt through more than 90 degrees
 at high luminosities. This suggest that radiation warping is a real phenomena.  
 
 There are two relevant timescales: the radiation torque timescale ($t_{\Gamma}$)
  and viscous timescale ($t_{\nu}$)
  \begin{equation}
t_{\Gamma} = \frac{12\pi \Sigma R^3 \Omega c}{L_{*}} \qquad {\rm and} \qquad  t_{\nu}=\frac{2R^2}{\nu}\ ,
\end{equation}
where $L_*$, is  the luminosity of the source; $R$, $\Sigma$,  $\Omega$, and  $\nu$, the radius, 
 the surface density, the Keplerian angular velocity and the viscosity  of the disk, respectively.

 For the inner ring,  the ring's surface density  is  $\Sigma \sim \rho_{\rm ring}  (2H_{\rm in})$ which gives
 \begin{equation}
t_{\Gamma} \sim  7.0 \ {\rm years}\    \frac{m_{\rm ring, -8} R_{\rm in, 25}^3 M_{QS, 1.5}^{1/2}}{R_{\rm out, 200}^{7/2} L_{\rm acc., 35}}  \ , 
\end{equation}
where the luminosity is given in units of $10^{35}$ erg s$^{-1}$.  The quark star mass, $M_{\rm QS}$, 
is given in units of $1.5M_{\odot}$.

 To estimate the viscosity we consider the physical state of the ring.
The ring is made up of solid degenerate walls (a few  meters thick) separated by thin layers of normal fluid.
The shear from the Keplerian angular velocity  is concentrated in the thin fluid layers.
Because of the inhibition of radial fluid motions (i.e. perpendicular to
the fluid layers) and thus inhibition of turbulence, we do not expect the magneto-rotational instability to develop in the thin fluid (this
remains to be confirmed).  The  thin fluid layers become the main contributor to the effective (overall) viscosity of the system.  As a result, we adopt  the standard Spitzer viscosity as an order of magnitude estimate.
The viscous timescale (see Appendix A in OLNII) is
\begin{equation}
\label{eq:vis}
  t_{\nu}\sim  21.9\ {\rm years}\ \frac{R_{\rm in, 25}^2}{(T_{\rm ring, keV}^{\rm b})^{5/2}}\ ,
\end{equation}
where the ring's temperature during the bursting phase (superscript ``b")  is in keV.

The $t_{\Gamma} < t_{\nu}$ condition (Eq. 4.1 in Pringle 1996) allows the
 instability to  develop and  sets the conditions for  retrograde motion.
 Thus,  ring warping occurs in our model when $L_{\rm acc.} > L_{\rm cr.}$ with 
\begin{eqnarray}
   L_{\rm cr.} &\sim& 3.2\times 10^{34}\ {\rm erg\ s}^{-1}\times \\\nonumber
   &&   \frac{m_{\rm ring, -8} R_{\rm in, 25} M_{QS, 1.5}^{1/2} (T_{\rm ring, keV}^{\rm b})^{5/2}}{R_{\rm out, 200}^{7/2}}\ .
\end{eqnarray}

 The 2002 outburst of AXP 1E2259$+$586 was above $3\times 10^{34}$ erg s$^{-1}$ for   hundreds of days (see Fig. 13 in Woods 2004
and Fig. 2 in OLNII).  
 For the instability to develop ($t_{\Gamma} < \sim $ 1 year; the length of time of $L> L_{\rm cr.}$), we require a reduction in $t_{\Gamma}$   by a factor of 10 or so  for our fiducial values.   Pringle (1996) notes
in \S 5 that a disk wind induced by the radiation would increase the effectiveness of momentum
transfer thus reducing $t_{\rm \Gamma}$ by a factor of $v_{\rm K, in}/c\sim 0.1$ where $v_{\rm K, in}$
is the Keplerian velocity at $R_{\rm in}$. 
  Furthermore, it is important to note
   that the early part of the burst was significantly brighter which also shortens the effective $t_{\Gamma}$
 by a factor of a few.   Effectively  a net reduction of $t_{\Gamma}$  to $t_{\rm \Gamma, eff.} < t_{\Gamma}/10$ is not unreasonable. The required outburst energy to reverse the inner ring is thus $E_{\rm burst}\sim t_{\rm \Gamma, eff} L_{\rm cr.} <   3\times 10^{41}~ {\rm ergs}$. In our model, this is
 provided by accretion from the ring's atmosphere   as it settles back to its sub-keV, equilibrium, temperature  (see \S 4.4 and Fig.2  in OLNII).

\section{The 2012 ``anti-glitch"}

 After the 2002 burst the ring cools back down to sub-keV temperatures ($\sim 0.2$ keV) which results in $t_{\nu} \sim 1.2\times 10^3$ years (see eq. \ref{eq:vis}).   During this period, the retrograde inner ring is penetrated by the magnetic field on timescales $\tau_{\rm B}\sim 881~ {\rm yrs}\times f_{\rm Fe}^2 R_{\rm in, 25}^3/\rho_{\rm ring, 3}^{1/6} $ (eq. 17 in OLNII). 
   The magnetic field penetration forces   the innermost retrograde wall of the ring to  co-rotate with the star. 
   This leads to  the accretion of the retrograde  wall and transfer of its angular momentum to  the star
  inducing  a sudden spin-down (i.e. ``anti-glitch").   To account for the 10 year interval  between the 2002 glitch and
  the 2012 anti-glitch (i.e. $\tau_{\rm B}\sim 10$ years) requires a dimensionless tensile strength $f_{\rm Fe}\sim 0.1$ in our model 
  which is not unreasonable (see \S 2.1 in OLNII).

   The  induced change in the star's frequency is 
   $\Delta \nu/\nu = - (M_{\rm wall} R_{\rm in}^2 \Omega_{\rm K, in})/(2/5 M_{\rm QS} R_{\rm QS}^2 \Omega_{\rm QS})$ (see
   eq. 29 in OLNII;  equation below gives the corrected version of eq. 30 in OLNII) which gives
\begin{equation}
\frac{\Delta \nu}{\nu} \sim - 6.1 \times 10^{-7} ~\frac{P_{\rm QS, 10} R_{\rm in, 25}^{1/2} M_{\rm wall, -11}}{M_{\rm QS, 1.5}^{1/2} R_{\rm QS, 10}^2}\ ,
\end{equation}
where $P_{\rm QS, 10}$,  $R_{\rm QS, 10}$ are the period ($P_{\rm QS}= 2\pi/\Omega_{\rm QS}$; in units of 10 s) and radius (in units of 10 km) of the quark star, respectively. The wall's mass is in units of $10^{-11}M_{\odot}$.  The above value is an upper limit: if the inner ring is not completely reversed,
 the angular momentum transfer is reduced by a factor $\cos{\theta}$ where $\theta$ is the tilt (i.e. the angle between the orbit normal and the equatorial  plane normal). Archibald et al. (2013)
 give $\Delta \nu/\nu \sim -3.1\times 10^{-7}$ to $\sim -6.3\times 10^{-7}$ using two different models which is
 consistent with our model estimates.  Archibald et al (2013) observed a period of increased spin-down following the ``anti-glitch" which can be explained in our model as described in \S 5.2 in OLNII.
     
 \subsection{The associated 2012 burst}    
     
  The accretion of  the   innermost  wall  occurs 
on timescales exceeding the free-fall time of a few milliseconds. The resulting 
   burst energy is $\sim \eta M_{\rm wall} c^2 = 0.1 M_{\rm wall}c^2\sim  10^{42}$ ergs.
 Archibald et al. (2013) give a luminosity of about $10^{38}$ erg s$^{-1}$ in the 10-1000 keV 
  band  over a time of  36 ms.  The discrepancy between the observed and predicted burst energy
   is not yet understood  but may be a consequence of a lower
   mass-to-radiation energy conversion efficiency $\eta$ ($<< 0.1$).   In the QN model, the QS is bare and crustless\footnote{This is the reason for the featureless spectrum emanating from the QS  in our model.} since it is in 
   the Color-Flavor Locked phase  which is 
    rigorously electrically neutral (Rajagopal \& Wilczek 2001); we assume that the surface depletion of strange quarks is negligible (see discussion in Usov 2004).
    Hadronic matter falling onto the QS will convert into strange quark matter releasing mostly neutrinos.
     This implies a reduced mass-to-radiation energy conversion efficiency factor ($\eta$) during accretion events.
     Combined with the fact that the wall is accreted as chunks (rather than fluid as in standard accretion) this may significantly 
    reduce heating and subsequent radiation during infall. Thus a reduction of  $\eta$ by a few orders of
    magnitudes is plausible which may help resolve  
     the discrepancy in burst energy in our model and the observed value.

\section{Model Limitations and  Predictions}

In summary, in this picture a bright burst ($\sim 10^{41}$ ergs) is required to reverse the inner ring. After several years, the reversed
 innermost wall  is accreted to cause the anti-glitch. 
Our model relies heavily on the assumption that radiation-induced warping can occur in the degenerate ring surrounding the quark star and that it would lead to the formation
 of a retrograde inner ring.  Only detailed numerical simulations (beyond the scope of this paper) can  prove or disprove this assumption. 
  The simulations could also include a more robust treatment of the viscosity of the ring. Such simulations could track the evolution of the retrograde ring  and  test   if  it could effectively remain stable for a few years  before it is accreted,  as seem to be the case in  AXP 1E2259$+$586. 
  In addition,  the burst energy in our model is a few orders of magnitude higher
 than the observed value. Unless the mass-to-radiation energy conversion efficiency $\eta$ is drastically reduced due to
  surface properties of the QS, this would be a major challenge to our model.\\

 Our model has the following features and predictions:

\begin{itemize}

\item If the first outburst (in this case the 2002 event) is bright enough to reverse more than one wall, 
 then we expect different outcomes. For example, more than one reversed wall
  can be accreted in an episode and also reversed and normal walls can be accreted in the same episode.
 Interestingly, Archibald et al. (2013)  mention two possibilities, an anti-glitch-anti-glitch pair
 or an anti-glitch-glitch pair fits their data.

 \item  If the anti-glitch is bright enough,  it can also lead to a reversal of  the 
  next inner ring. In this case an ``anti-glitch"  would be associated with the subsequent
   outburst (roughly 10 years later).

   \item   We expect a possible outburst  within the next decade (around year $\sim$ 2022), since  the time it takes
   the magnetic field to penetrate the counter-rotating innermost wall and accrete it is
   $\tau_{\rm B}\sim 8.81~ {\rm yrs}\times f_{\rm Fe, 0.1}^2 R_{\rm in, 25}^3/\rho_{\rm ring, 3}^{1/6} $.

\item  The fall-back material  is representative of the QN ejecta and is thus rich
   in heavy elements (Jaikumar et al. 2007).    If heated to keV temperatures 
   we expect atomic lines to be detected (e.g. Koning et al. 2013).

\end{itemize}

\begin{acknowledgements}

This work is funded by the Natural Sciences and Engineering Research Council of Canada. N.K. would like to acknowledge support from the Killam Trusts.

\end{acknowledgements}


\end{document}